\documentclass{ifacconf}

\makeatletter
\let\old@ssect\@ssect 
\makeatother

\usepackage{graphicx}      
\usepackage[square,numbers]{natbib}        

\usepackage[colorlinks=false]{hyperref}

\makeatletter
\def\@ssect#1#2#3#4#5#6{%
  \NR@gettitle{#6}
  \old@ssect{#1}{#2}{#3}{#4}{#5}{#6}
}
\makeatother

\usepackage{subfigure}

\usepackage{xcolor}

\newcommand{\gc}[1]{{\color{blue}[Gin: #1]}}

\usepackage{cancel}

\let\theoremstyle\relax
\usepackage{amsmath,amssymb,amsthm}
\usepackage{mathrsfs}
\usepackage{physics}

\usepackage{algorithm}
\usepackage[noend]{algpseudocode}

\usepackage{booktabs}

\begin{document}
\begin{frontmatter}

\title{Adversarial Learning of Robust and Safe Controllers for Cyber-Physical Systems} 

\thanks[footnoteinfo]{This work has been partially supported by the PRIN project “SEDUCE” n. 2017TWRCNB.}

\author[First,Second]{Luca Bortolussi} 
\author[First]{Francesca Cairoli} 
\author[First]{Ginevra Carbone}
\author[First]{Francesco Franchina}
\author[First]{Enrico Regolin}

\address[First]{University of Trieste, Italy (e-mail: lbortolussi@units.it - francesca.cairoli@phd.units.it - ginevra.carbone@phd.units.it - francesco.franchina@studenti.units.it - eregolin@units.it).}
\address[Second]{Saarland University, Saarbr\"ucken, Germany.}

\begin{abstract}                
We introduce a novel learning-based approach to synthesize safe and robust controllers for autonomous Cyber-Physical Systems and, at the same time, to generate challenging tests.
This procedure combines formal methods for model verification with Generative Adversarial Networks. The method learns two Neural Networks: the first one aims at generating troubling scenarios for the controller, while the second one aims at enforcing the safety constraints. We test the proposed method on a variety of case studies.
\end{abstract}


\end{frontmatter}

\section{Introduction}

Controlling Cyber-Physical Systems (CPS) is a well-established problem in classic control theory ~\citep{pidrulez}. State of the art solutions apply to all those models in which a complete knowledge of the system is available, i.e., scenarios in which the environment is supposed to follow deterministic rules. For such models a high level of predictability, along with good robustness, is achieved.
However, as soon as these unpredictable scenarios come into play, traditional controllers 
are challenged and could fail.
Ongoing research is trying to guarantee more flexibility and resilience in this context by using Deep Learning ~\citep{deepqlearning} and, in particular, Reinforcement Learning for robust control. 
State of the art solutions perform reasonably well, but they still present evident limits in case of unexpected situations. The so called \textit{open world scenarios} are difficult to
model and to control, due to the significant amount of stochastic variables that are needed in their modelling and to the variety of uncertain scenarios that they present. Therefore, while trying to ensure safety and robustness, we need to be cautious about not trading them with model effectiveness.

In this work we investigate autonomous learning of safe and robust controllers in open world scenarios.
Our approach consists in training two neural networks, inspired by Generative Adversarial Networks (GAN)~\citep{goodfellow2014generative}, that have opposite goals: the \textit{attacker} network tries to generate troubling scenarios for the \textit{defender}, which in turn tries to learn how to face them without violating some safety constraints. The outcome of this training procedure is twofold: on the one hand we get a robust controller, whereas on the other hand we get a generator of adverse tests.\footnote{Code is available at: \url{https://github.com/ginevracoal/adversarialGAN/}}

The learned controller is a black-box device capable of dealing with adverse or unobserved scenarios, though without any worst-case guarantee. In this regard, one could complement our method with a shield-based approach, as proposed e.g. by~\cite{avni2019run}.

\section{Problem Statement and Related Work}

Safety of a system can be formalised as the satisfaction of a set of safety requirements. 
A popular approach to mathematically express such safety requirements is \emph{Signal Temporal Logic} (STL) \citep{robust_stl_2010}. 
Temporal logic is used in the context of \emph{formal verification} to express the desired behaviour of a system in terms of time. It extends propositional logic with a set of modal operators capturing the temporal relation among events \citep{sep-logic-temporal}. STL formulas, in particular, deal with properties of continuous-time real-valued signals, such as CPS trajectories.
In our application, we express safety requirements only by means of time-bounded formulas over fixed-length trajectories. In particular, we rely on STL \emph{quantitative semantics}, which is capable of capturing, for each trajectory, the level of satisfaction of the desired property by measuring how much the input trajectory can be shifted without changing its truth value. Such measure is often referred to as \emph{robustness} and it is exploited in this work as the objective function of an optimization problem. 


We model the interaction of an agent with an adversarial environment as a \emph{zero-sum game}, similarly to the strategy behind GANs. 
The concept of zero-sum game is borrowed from game theory and denotes those situations in which one player's gain is equivalent to another's loss. In such situations, the best strategy for each player is to minimize its loss, while assuming that the opponent is playing at its best. This concept is known in literature as \textit{minmax strategy}. 
In practice, we use GAN architectural and theoretical design to reach two main objectives: a controller, that safely acts under adverse conditions, and an attacker, which gains insights about troubling scenarios for the opponent. The concept is closely related to that of \emph{Robust Adversarial Reinforcement Learning (RARL)} ~\citep{pinto2017robust}, a Reinforcement Learning framework, involving an agent and a destabilizing opponent, that is robust to adverse environmental disturbances.
In this case the term ``robustness'' does not refer to Signal Temporal Logic, but instead, to the cumulative reward computed on the learned policy with respect to the varying test conditions. {Other recent RL techniques involving STL constraints include \cite{ balakrishnan2019structured, bozkurt2020model,liu2021recurrent}}
\section{Methodology}

\emph{Agent-Environment Model.}
Due to coexistence of continuous and discrete components, CPSs are typically represented as hybrid models: the continuous part is represented by differential equations that describe the behaviour of the plant; the discrete part, instead, identifies the possible states of the controller. 
We decompose our model in two interacting parts: the \emph{agent} $a$ and the \emph{environment} $e$. Both of them are able to observe at least part of the whole state space $\mathcal{S}$, i.e. they are aware of some observable states $\mathcal{O}\subset\mathcal{S}$.
By distinguishing between the observable states of the agent $\mathcal{O}_a\subseteq \mathcal{O}$ and of the environment $\mathcal{O}_e\subseteq \mathcal{O}$, we are able to force uneven levels of knowledge between them. 
Notice that the observable states, for both the agent and the environment, could also include environmental variables involved in the evolution of the system.

Let $\mathcal{U}_a$ and $\mathcal{U}_e$ be the spaces of all possible actions for the two components. We discretize the evolution of the system as a discrete-time system with step $\Delta t$, which evolves according to a function $\psi: \mathcal{S} \times \mathcal{U}_a \times \mathcal{U}_e \times \mathbb{R} \longrightarrow \mathcal{S}$.
By taking control actions at fixed time intervals of length $\Delta t$, we obtain a discrete evolution of the form  $ s_{i+1} = s_i + \psi(s_i, u_a^i, u_e^i, t_i)$, where $t_i := t_0 + i\cdot\Delta t$, $u^i := u(t_i)$ and $s_i := s(t_i)$.  Therefore, we are able to simulate the entire evolution of the system over a time horizon $H$ via $\psi$ and to obtain a discrete trajectory $\xi=s_0 \dots s_{H-1}$. 

\emph{STL syntax.} The STL syntax is defined by
$$
\varphi := \mathtt{true} \mid f (s) > 0 \mid \neg\varphi \mid \varphi_1 \land \varphi_2 \mid \varphi_1\, \mathtt{U}_{[a,b]} \varphi_2
$$
where $s : \mathbb{R}^+\rightarrow\mathcal{S}$
is a signal, $f : \mathcal{S}\rightarrow\mathbb{R}$ is a real-valued function, and $[a,b]$ is an interval of non-negative real numbers in the time domain $\mathbb{R}^+$. Two important temporal operator can be derived from the syntax above: the \textit{eventually} operator $\Diamond_{[a,b]} \varphi \equiv \mathtt{true}\, \mathtt{U}_{[a,b]} \varphi$, and the \textit{globally} operator $\square_{[a,b]} \varphi\equiv\neg (\Diamond_{[a,b]} \neg\varphi)$.
The definition of Boolean and quantitative semantics is given in Section~\ref{sec:semantics} of the Appendix. 

A safety requirement is expressed as an STL formula $\varphi$; we call $h$ its temporal depth.\footnote{The temporal depth of a formula is defined recursively as the sum of maximum bounds of nested temporal operators.}
The robustness of a trajectory quantifies the level of satisfaction w.r.t. $\varphi$ and it determines how safe the system is in that configuration. Robustness is denoted 
as a function $R_{\varphi}: \mathcal{S}^h \rightarrow \mathbb{R}$, measuring the maximum perturbation that can be applied to a given trajectory of length $h$ without changing its truth value w.r.t. $\varphi$. 
It is straightforward to use this measure as the objective function in our minmax game. 

\emph{Multi-objective formulation.}
In case of multiple safety requirements, we define a different STL formula for each of these requirements and likewise we compute the respective robustness values. As a matter of fact, the order of magnitude of each robustness value depends on the order of magnitude of the CPS variables involved. Therefore, a single STL formula that combines the safety requirements all together may result in an unbalanced objective function, skewed towards some components that are not necessarily the most safety-critical. To overcome this problem we normalize the variables involved and we define the objective function as a weighted sum of the robustness values $R_{\varphi_i}$ resulting from each requirement $\varphi_i$. 
Let $\Phi=\{\varphi_1,\dots, \varphi_m\}$ denote a set of $m$ safety requirements, the combined robustness score $R_{\Phi}$ is defined as:
\begin{equation}
\label{eqn:robustness_score}
R_{\Phi}(\cdot) := \frac{1}{\alpha}\sum_{i=1}^m \alpha_i\cdot R_{\varphi_i}(\cdot),    
\end{equation}
where $\alpha= \sum_{i=1}^m \alpha_i$.
By tuning the weights $\alpha_1,\dots, \alpha_m$ we are able to explicitly influence the importance of each factor in the training objective.  The choice of these hyper-parameters will be case-specific. This formulation is typically used in multi-objective optimization scenarios ~\citep{li_belta}. 
In \eqref{eqn:robustness_score}, we assume w.l.o.g. for notational simplicity that each formula $\varphi_i$ has the same time depth $h$.
However, our framework can be straightforwardly extended to more general STL properties with different time depths.   
\emph{Attacker-Defender architecture.} 
The proposed framework builds on GAN architectural design, in which two NNs compete in a minmax game to reach opposite goals.
One network, denoted by $A$, represents the \textit{attacker}, while the other, denoted by $D$, represents the \textit{defender}. The aim of the former is to generate environment configurations in which the defender is not able to act safely, whereas, the latter tries to keep the CPS as safe as possible. 
In practice, the defender $D$ can be interpreted as a controller for the agent.

\emph{Optimization strategy.}
Given a time horizon $H$, an initial state $s_0$, meaning the state at time $t_0$, and two sequences of actions $\textbf{u}_a = (u_a^0, \dots, u_a^{H-1} )$ and $ \textbf{u}_e = (u_e^0, \dots, u_e^{H-1} )$, one for the agent and one for the environment, it follows that the evolution of a trajectory $\xi$ is obtained by
evaluating $\psi$ at each time steps $t_i\in\{t_0,\dots, t_{H-1}\}$.
The minmax problem can be expressed as finding the sequences $\textbf{u}_e$ and $\textbf{u}_a$ that solve
\begin{equation}
\min_{\textbf{u}_e}\max_{\textbf{u}_a} \left[ \mathcal{J}(s_0, \textbf{u}_a, \textbf{u}_e)\right].
\end{equation}
The objective function $\mathcal{J}$ is the cumulative sum of the robustness scores computed at each timestep during the generation of the whole trajectory $\xi$ on the sub-trajectory $\xi[t,t+h-1]$ available at timestep $t$, i.e.
$$
\mathcal{J}(s_0, \textbf{u}_a, \textbf{u}_e) := 
\sum_{t=0}^{H-h}R_{\Phi}(\xi[t,t+h-1]).
$$

In our setting, the sequences of actions are iteratively determined by the two adversarial networks. In particular, let $\theta_A$ be the weights of the attacker's network $A$ and $\theta_D$ the weights of the defender's network $D$. At each timestep, the attacking network, 
\begin{align*}
A: \Theta_A \times \mathcal{O}_e \times \mathcal{Z} &\longrightarrow \mathcal{U}_e \\
(\theta_A, \textbf{o}_e, \textbf{z}) & \longmapsto u_e,
\end{align*} 
receives the current observable state of the environment $\mathbf{o}_e$, the noise coefficient $\mathbf{z}$ and outputs the coefficients $u_e$, defining the adversarial environmental components. 
Similarly, the defender network,
\begin{align*}
D: \Theta_D \times \mathcal{O}_a &\longrightarrow \mathcal{U}_a \\
(\theta_D, \textbf{o}_a) & \longmapsto u_a,
\end{align*}
reads the current observable state of the agent $\mathbf{o}_a$ and produces the control action $u_a$.

To ease the notation, we introduce a function
\begin{align*}
\psi_t: \mathcal{S} \times \Theta_D \times \Theta_A &\longrightarrow \mathcal{S}^t
\\
(s_0, \theta_D, \theta_A) &\longmapsto s_0 \dots s_{t-1},
\end{align*}
which iteratively applies $\psi$ for each pair of actions 
\begin{align*}
u_a^j &= D(\theta_D, \textbf{o}_a^j)\\
u_e^j &= A(\theta_A, \textbf{o}_e^j, \textbf{z}), 
\end{align*}
where $j\in\{0,\ldots, t-1\}$ indexes the simulation interval.

The formalism introduced by the two policy networks transforms the
problem of finding the best sequences of actions, $\textbf{u}_a$ and $\textbf{u}_e$, to that of finding the best networks' parameters, $\theta_D$ and $\theta_A$.
This leads to the objective 
\begin{equation}
J(s_0, \theta_A, \theta_D) =
\sum_{t=0}^{H-h}
R_{\Phi}[\psi_h(s_t, \theta_D, \theta_A)]
\label{eq:objective}
\end{equation}
and the minmax game $ {\min}_{\theta_A} \; {\max}_{\theta_D} \; J(s_0, \theta_A, \theta_D)$ is now directly expressed in terms of the training parameters. 

In such setting, the defender aims at generating safe actions by tuning its weights in favour of a maximization of the objective function, i.e., a maximization of the cumulative robustness score. The attacker, on the other hand, aims at generating troubling scenarios for the opponent by minimizing the objective function, i.e., minimizing the cumulative robustness score.

The horizon $H$ represents the number of simulation steps performed while keeping the parameters $\theta_A$ and $\theta_D$ fixed.
In principle, we could choose $H=h$, without the need of having a summation in \eqref{eq:objective}, possibly taking a larger time bound $h$ in the formulae of $\Phi$. However, this would make the objective excessively rigid. In fact, if a controller would work well everywhere but on a small sub-region of the trajectory, such an objective would return a penalization also for the regions where the controller performs well. This effect is avoided by considering $h\ll H$ in the objective \eqref{eq:objective}. 



\emph{Testing phase.} In the testing phase, we generate a trajectory of length $H$ and we check separately each safety requirement on such trajectory, in particular we check that the requirement $\varphi_i$ is globally satisfied, i.e., the condition $\square_{[0,H]} \varphi_i$. 
Therefore, for each property, a positive value of robustness at test time means that the requirement is met during the whole evolution of the system w.r.t. the time horizon $H$.
The training and testing pseudo-codes are shown in Section \ref{sec:pseudocode} of the Appendix.

\section{Experiments}
We test the proposed architecture on two different case studies: a cart-pole balancing problem and a platooning problem. Both systems are embedded into environments with a stochastic evolution. 

The Attacker-Defender networks are trained against each other for a given number of epochs. Once the training is over, the performances of the trained Defender are tested in two different ways. On one hand, we generate a test set containing 1k different initial configurations, uniformly sampled from pre-defined compact sets (see Tab. \ref{tab:cartpole_params} and \ref{tab:platooning_params} in the Appendix). 
From each of these points we generate a trajectory evolving according to the trained Attacker and Defender networks, then check each requirement separately on each trajectory as specified in the previous section.
The second approach to evaluate the performances of the trained Attacker-Defender network is to consider an environment that evolves unaware of the state of the system. In both cases, we compare the performance of Defender with that of a classical controller.

Hyperparameter tuning is necessary for the GAN architecture to achieve the desired performances in terms of safety. The choice of the architecture (number and size of the layers), the training hyperparameters, the time horizon $H$ and weights for the cumulative robustness have a strong impact on the final results. In  particular, the number of training iterations performed by the Attacker network and by the Defender network has a strong impact on the performances of the trained networks. By tuning this number we are able to ensure that the Attacker is strong enough to generate challenging configurations of the environment, without preventing the Defender network from learning a secure controller. {We performed manual tuning on a combination of hyperparameters and architectures, however one could also automate this process by maximizing the percentage of safe trajectories produced by the learned controller.}

\subsection{Cart-Pole balancing}
\label{sec:cartpole}

The \emph{Cart-Pole} system ~\citep{florian} (also known as Inverted Pendulum) consists of a cart and a vertical pole attached to the cart by an unactuated joint. The cart is allowed to move along the horizontal axis, while the pole moves is the vertical plane parallel to the track. 
The goal is to keep the pole balanced by learning an optimal policy for the cart, which influences the swinging movement of the pole. This problem is a well known benchmark in both classical control ~\citep{aguilar2014stabilization, liu2008tracking} and reinforcement learning  ~\citep{nagendra2017comparison, lillicrap2019continuous} applications.

\emph{Moving target and track-cart friction.} In order to test the full potential of our framework, we consider a complex stochastic environment made of a moving target for the cart to follow and a friction coefficient between the cart and its track. These two components, governed by the Attacker network, represent the two potentially adversarial components of the system. 

\emph{Model.} The observable states $\mathbf{o}_a$ for the Defender and $\mathbf{o}_e$ for the Attacker are: cart position $x$, cart velocity $\dot x$, pole angle $\theta$, pole angular velocity $\dot \theta$ and target position $\hat{x}$. 
Given 
$\mathbf{o}_e$, the Attacker's policy network $A$ generates the adverse coefficients, i.e., friction $\mu$ and target position $\hat{x}$, 
both constrained to assume realistic values w.r.t. the physical settings of our application.
The Defender reads the current state $\mathbf{o}_a$ and generates the desired control action $f$ for the cart, which is meant to keep the pole balanced during the whole trajectory. 
The dynamic of the system is described by the following equations~\citep{cartpolewang}:
\begin{equation}
\begin{cases}
    \ddot x &= \frac{f - \mu \dot x + m_p\,l\,\dot\theta^2\,\sin\theta-m_p\,g\,\cos\theta\sin\theta}{m_c+m_p\,\sin\theta^2}
\\
    \ddot\theta&=\frac{g\,\sin\theta-\cos\theta\ddot x}{l}
\end{cases},
\end{equation}
where $m_p$ is the mass of the pole, $m_c$ is the mass of the cart, $l$ is half the pole length and $g$ is the gravitational constant.

We impose the STL requirement $\varphi_d = \square_{[0,h]}(d \leq d_{\text{max}} \wedge d \geq d_{\text{min}})$ on the distance $d=\|x-\hat{x}\|$ between the cart and its target, where $d_{\text{min}}$ and $d_{\text{max}}$ are the minimum and maximum distances allowed. Similarly, we set the requirement $\varphi_\theta = \square_{[0,h]}(\theta \leq \theta_{\text{max}} \wedge \theta \geq \theta_{\text{min}})$ on the angle.

The objective function is the combination of two cumulative robustness components, one on the distance, $R_{\varphi_d}$, and one on the angle, $R_{\varphi_\theta}$, whose contributions are weighted by a coefficient $\alpha\in[0,1]$:
$$
J(s_0, \theta_A, \theta_D) = \alpha\,\sum_{t=0}^{H-h}R_{\varphi_d}[\xi_t] + (1-\alpha) \,\sum_{t=0}^{H-h}R_{\varphi_\theta}[\xi_t],
$$
where $\xi_t = \psi_h(s_t,\theta_D,\theta_A)$.

\begin{figure}[!t]
\begin{subfigure}
    \centering
	\includegraphics[width=9cm, keepaspectratio]{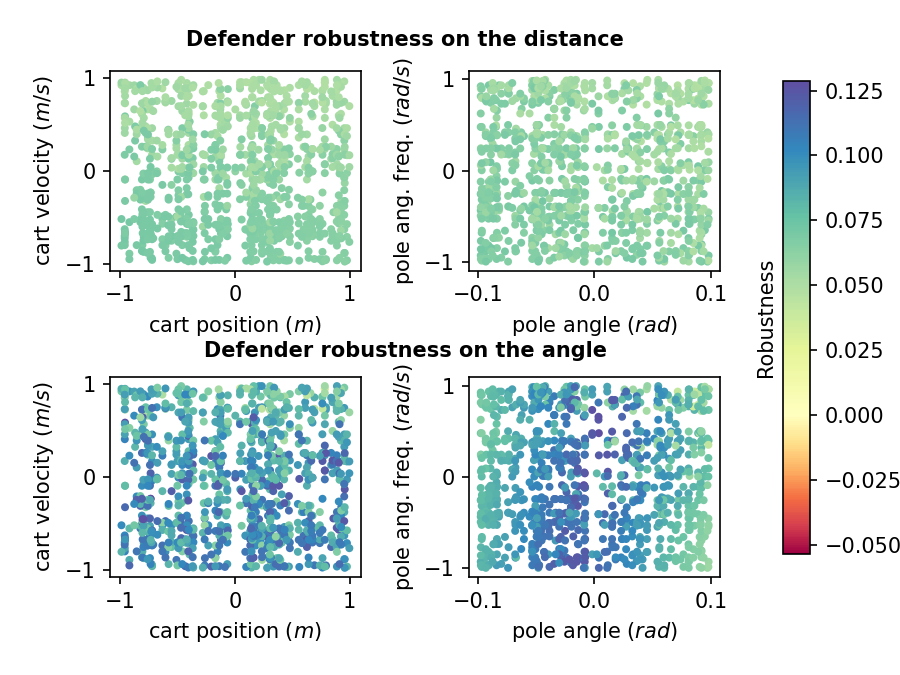}
\end{subfigure}
\begin{subfigure}
    \centering
	\includegraphics[width=9cm, keepaspectratio]{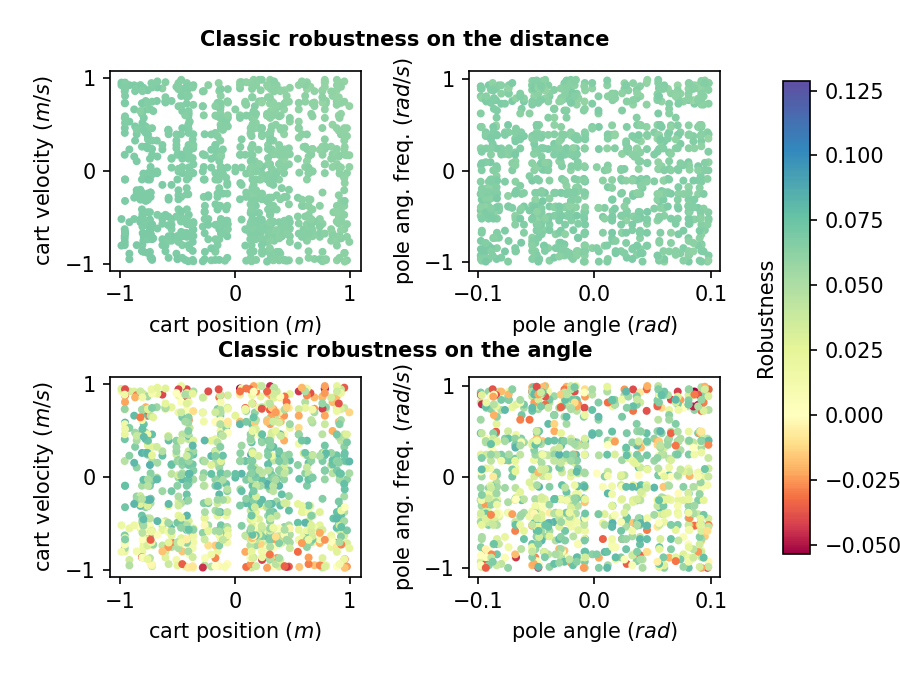}
\end{subfigure}
\caption{Cartpole balancing problem in which the friction and the moving target are generated by the Attacker network. This plot shows the robustness values achieved by the Defender and the classic controller. Robustness differences are computed separately for the two requirements imposed on the distance and on the angle. Trajectories start from 1k random initial states and evolve on a time horizon $H=200$ with a step of $\Delta t=0.05\,s$, resulting in $\Delta t \cdot H=10\;s$ long simulations. 
}
\label{fig:cartpole_atk_scatterplot}
\end{figure}

\begin{figure}[!t]
    \centering
	\includegraphics[width=9cm, keepaspectratio]{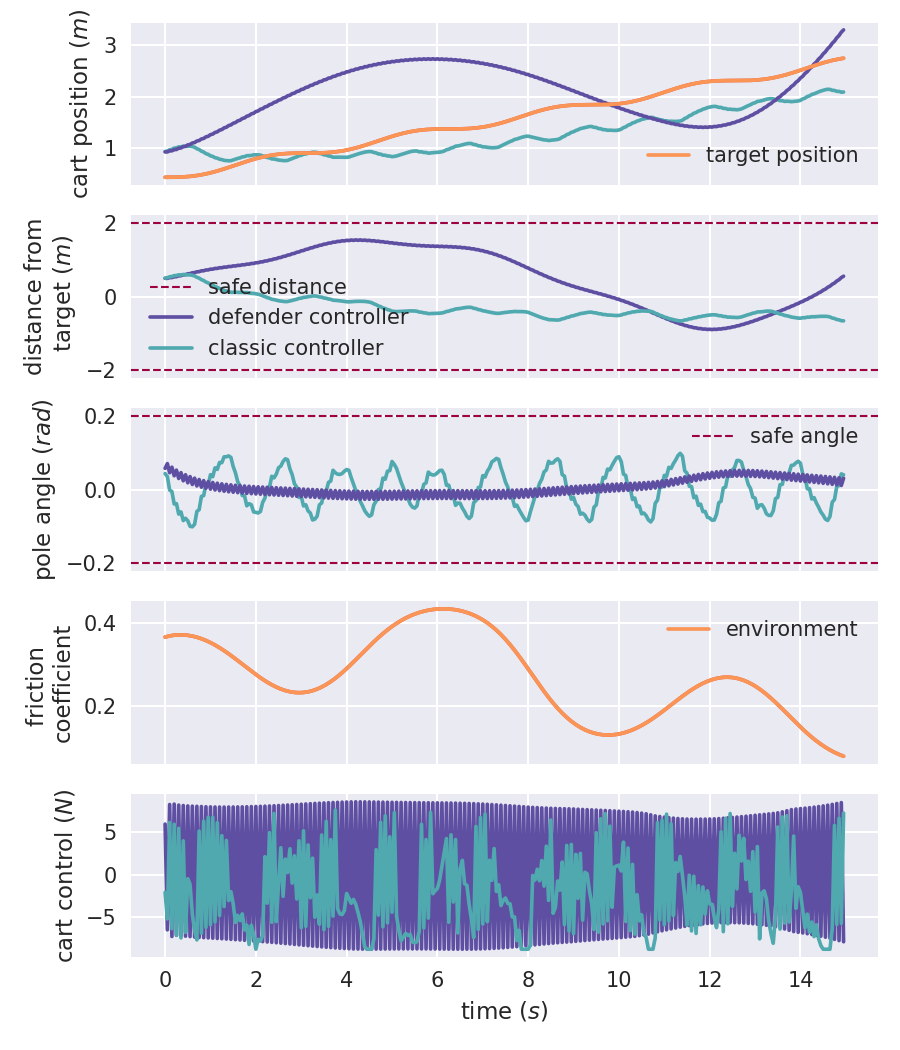}
	\caption{Sample of the system evolution for the cartpole balancing problem, using a classical controller and the Defender network against the same environment, represented by fixed trajectories of target positions and the cart-track friction coefficients.  The \emph{defender} cart is controlled by the Defender network, while the \emph{classic} one is managed a classical controller.  The common initial state 
	is: cart position ${x=0.9055}\;m$, cart velocity ${\dot x=0.0348}\;m/s$, pole angle ${\theta=0.0808}\;rad$, pole angular velocity ${\dot\theta=0.0399}\;rad/s$. The trajectory evolves on a time horizon $H=300$ with time step size $\Delta t=0.05\;s$ ($\Delta t \cdot H=15\;s$ long simulation).}
	\label{fig:cartpole_pulse_evolution}
\end{figure}

\emph{Results.}
The experimental settings are presented in Table \ref{tab:cartpole_params} of the Appendix. 
Fig. \ref{fig:cartpole_atk_scatterplot} shows that the
trajectories evolving according to the Defender network all achieved positive robustness, despite the adversarial reactive environment governed by the Attacker. Moreover, Fig. \ref{fig:cartpole_pulse_evolution} shows the evolution of the system in a fixed environmental setting, for two different controllers: the Defender network and a classical robust controller based on a Sliding Mode Control (SMC) architecture \citep{edwards1998sliding}.
The Defender is able to maintain safety during the whole trajectory on both $\theta$ and $d$, adequately counteracting the cart-track friction, while the classical controller has a worse overall performance and
in a few cases failed to guarantee safety within the specified initialization grid.
In this setting the Defender exhibits chattering in its control signal. In fast evolving systems, such as cartpole, this phenomenon could be avoided by including a regularization term on the control signal during the training phase. It should be noticed that the relatively low sampling frequency of $20Hz$ could also affect the SMC based controller. 

\subsection{Car platooning} 
\label{sec:platooning}
A \emph{platoon} ~\citep{carplatooning} is a group of vehicles travelling together very closely and safely. This problem is usually faced with techniques that coordinate the actions of the entire pool of vehicles as a single entity
~\citep{platooning_survey}. This approach, though, requires specific hardware and a distributed system of coordination that might be difficult to realise in complex scenarios. Our method, instead, builds a robust controller for individual decision-making, hence it fits into the autonomous driving field. In this setting, we assume that all vehicles are equipped with an hardware component called \emph{LIDAR scanner}, which is able to measure the distance between two cars by using a laser beam. 
In the basic scenario, only involving two cars, the car in front is called the \emph{leader} and acts according to the Attacker network, while the second one is the \emph{follower}, whose behaviour is determined by the Defender network. This setting trivially extends to the case of $n$ cars, where the first car is the leader and the other ones all act as followers, controlled by the same Defender.

\emph{Platooning with Power Consumption}

An additional problem that can be addressed in the platooning problem is the optimization of the energy consumption of the follower car, similarly to what has been proposed by \citep{zambelli2019robustified}, who exploited a non-cooperative distributed MPC framework. 

We are given a vehicle with mass $m$ and effective wheel radius $R_e$, which moves on a flat straight line covering the distance $x(t)$, with the driver inputs producing a torque at wheel level $T_{w}$. We factor the loss in two terms, which include the effects of rolling resistance (deformation of the rolling wheel on the ground) and aerodynamic resistance.
The dynamics follows the following equation
\begin{equation}
m\ddot{x}(t) = \frac{T_{w}(t)}{R_e} - C_r m g \dot{x}(t) -  \frac{1}{2}\rho C_a S \dot{x}^2(t),
\end{equation}
where $C_r$ is the rolling resistance coefficient,  $g$ the gravitational acceleration, $\rho$ the air density, $C_a$ the aerodynamic coefficient and $S$ the equivalent vehicle surface.

The torque at wheel level $T_{w}(t)$ is a function of time given by the combined effect of electrical torque at motor level, $T_{m}$, and the one due to the conventional brake action, $T_b$, that acts directly on the brake calipers. By factoring in the gear ratio between motor and wheels, one has:
\begin{equation}
T_{w}(t) = T_{m}(t)\cdot r_g + T_b(t).
\end{equation}

From an energy efficiency standpoint, it is reasonable to expect an optimal platooning policy to minimize the overall consumed electrical energy. This can be achieved by operating as much as possible the electric powertrain in its most efficient working point, and by avoiding situations in which the conventional brake has to be operated for safety reasons, i.e. when the safety requirement on the distance is violated. 

With regards to powertrain efficiency, we define an efficiency map $\eta(T_{m}(t),\omega_m(t))$ 
which combines the effects of battery and e-motor, where $\omega_m$ is the motor speed. At time $t$ the consumed electric power is  
\begin{equation}
P_m(t) = T_{m}(t)\omega_m(t)\eta(T_{m}(t),\omega_m(t))^{-sign(T_m(t))}.
\end{equation}

In a single-gear setting, the motor speed is related to vehicle speed through the relation $\omega_m(t) = \frac{r_g}{R_e}\dot{x}(t)$.


\emph{Model.}
We  consider the simple case of two cars, one \emph{leader} $l$ and one \emph{follower} $f$, whose internal states are position $x$, velocity $v$ and acceleration $a$.
The follower $f$ acts as the agent of this system, while the leader $l$ is considered to be part of the adversarial environment, to simulate a cyberattack scenario.
They have the same observable states $\textbf{o}_a=\textbf{o}_e=(\dot x_l, \dot x_f, d)$, given by their velocities and by their relative distance $d$. In the basic platooning setting the policy networks $A$ and $D$ output the accelerations $\textbf{u}_a=\ddot x_f$ and $\textbf{u}_e=\ddot x_l$,
which are used to update the internal states of both cars.
When the energy consumption evaluation is involved, the policy networks output electric and conventional torque values for the two cars, $\mathbf{u}_a=(T_{el}^a,T_b^a), \mathbf{u}_e=(T_{el}^e,T_b^e)$, that are used to compute the corresponding accelerations.
The dynamic of a car with mass $m$ and velocity $v$ is described as
$m \dv{v}{t} = m a_\text{in}  - \nu m g$,
where $a_\text{in}$ is the input acceleration produced by one of the two policies, $\nu$ is the friction coefficient and $g$ is the gravity constant. We impose the following STL requirements:  $\varphi_d = \square_{[0,h]}(d \leq d_{\text{max}} \wedge d \geq d_{\text{min}})$ on the distance $d=x_l-x_f$ between the two vehicles and $\varphi_e = \square_{[0,h]}(e \leq e_{\text{max}})$ on the power energy consumption $e$ (note that the higher the robustness of $\varphi_e$, the lower the energy consumption). 
The objective of our optimization problem is

$$ 
J(s_0, \theta_A, \theta_D) =\alpha\,\sum_{t=k}^{H-1}R_{\varphi_d}[\xi_t]+(1-\alpha)\,\sum_{t=k}^{H-1}R_{\varphi_e}[\xi_t],
$$
where $\xi_t = F_h(s_t,\theta_D,\theta_A)$, $R_{\varphi_d}$ is the robustness on the distance $d$ and $R_{\varphi_e}$ is the robustness of the energy consumption.
The extension to a platoon of $n$ cars is straightforward: the first car is the leader and each of the other cars follows the one in front. The first pair of subsequent cars acts as described in the two-cars model, while the other followers are controlled by copies of the same Defender network.

\begin{figure}[!t]
    \centering
	\includegraphics[width=8.5cm, keepaspectratio]{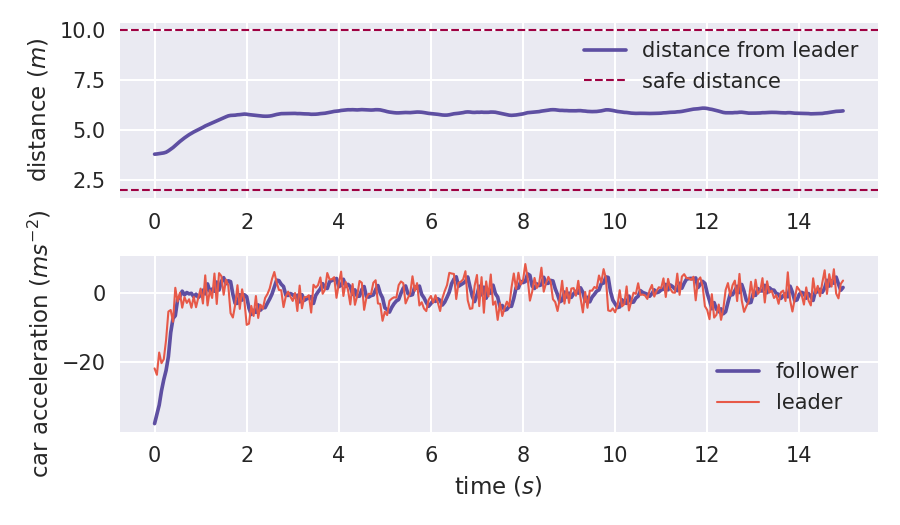}
	\caption{Defender network (follower) against Attacker network (leader) for car platooning problem. Initial state: distance between cars $d=4.2439\;m$, leader velocity $v_l=18.1229\;m/s$, follower velocity $v_f=15.7211\;m/s$.	The time horizon is $H=300$ and the step width is $\Delta t=0.05\,s$. 
	}
	\label{fig:platooning_atk}
\end{figure}

\begin{figure}[!t]
    \centering
	\includegraphics[width=8.5cm, keepaspectratio]{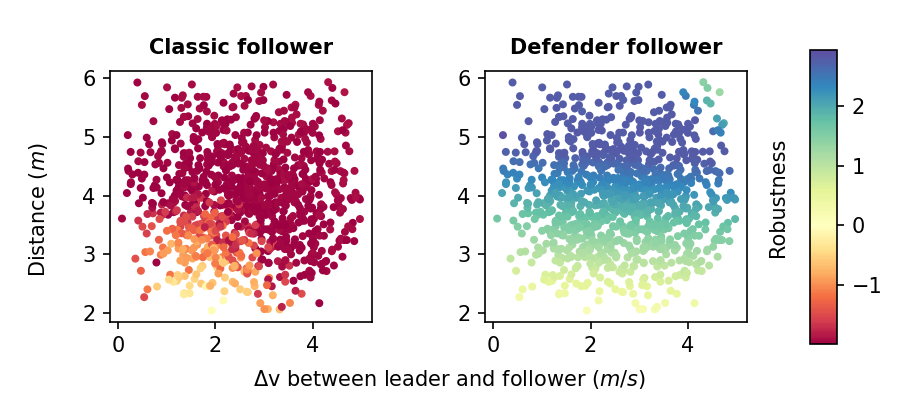}
	\caption{Global robustness values on the distance requirement for a classic follower and the Defender follower against the Attacker leader, starting from $1k$ different random configurations of the system (leader and follower cars positions and velocities). Random  initializations  of the  states  are described in Tab. \ref{tab:platooning_params} of the appendix.
	The time horizon is $H=200$ and the step width is $\Delta t=0.05\,s$. 
c	}
	\label{fig:platooning_atk_scatter}
\end{figure}

\emph{Results.}
In the basic platooning scenario, i.e. ignoring energy consumption, the leader acts according to the Attacker's policy, with sudden accelerations and brakes.
In such case, the follower learns to manage the unpredictable behaviour of the attacker by maintaining their relative distances within the safety range, as shown in Fig. \ref{fig:platooning_atk}.

Introducing requirements on the energy consumption makes the problem of platooning more challenging. Nonetheless, our architecture is still able to provide a safe controller. Fig. \ref{fig:platooning_pulse_evolution} in Sec. \ref{sec:additional_plots} of the Appendix shows an evolution of the system in which the leader car is unaware of the state of the follower and the two followers are managed by the Defender network and by a PID classical controller, which implements a basic  \lq\lq energy aware'' distance-tracking strategy. In this scenario, both controllers are able to ensure safety during the whole trajectory, although at the moment the defender appears to be focusing mostly on keeping the ideal distance from the leader, even applying an energy-inefficient strategy. 

However, when the leader actions are adverse reactions to the state of the follower, meaning actions chosen by the Attacker, the global robustness achieved by the Defender is in general much higher than the one achieved by the classic controller, as shown in Fig.\ref{fig:platooning_atk_scatter}.

\section{Conclusions}

Classical control theory struggles in giving adequate safety guarantees in many complex real world scenarios. New reinforcement learning techniques aim at modelling the behaviour of complex systems and learning optimal controllers from the observed data. Therefore, they are particularly suitable for stochastic optimal control problems where the transition dynamics and the reward functions are unknown. 
We proposed a new learning technique, whose architecture is inspired by Generative Adversarial Networks, and tested its full potential against the vehicle platooning and the cart-pole problem with additional stochastic components. {We tested the learned controllers against black-box adversarial policies in the case of completely observable systems, but our framework could also be straightforwardly extended to partially observable systems.} Our approach has been able to enforce safety of the model, while also gaining insights about adverse configurations of the environment.
As future work, we plan to improve the performances in the multi-objective case, to  test more complex scenarios, to investigate the scalability of this approach, and to introduce a regularization in the objective function in order to make the Defender policy less stiff, removing chattering behaviour.


\bibliography{bibliography}



\appendix

\section{Training and testing pseudocodes}
\label{sec:pseudocode}

\begin{algorithm}[H]
\begin{algorithmic}[1]
\Procedure{Train}{$\mathcal{M}$, $H_\text{train}$, $\text{iters}_A$, $\text{iters}_D$}
    \State $s_0 \gets$ SampleRandomState()
    \\
    \For{$\text{iters}_A$} \Comment{Train $A$}
        \State $\mathbf{u}_e$=[\:]
        \State $\mathbf{u}_a$=[\:]
        \\
        \For{$i \gets 0 \dots H_\text{train}-1$}   
            \State $\textbf{z} \gets \mathcal{N}(0,1)$
            \State $u_e^i \gets A(\theta_A, \textbf{o}_e^i, \textbf{z})$
            \State $u_a^i \gets D(\theta_D, \textbf{o}_a^i)$
            \State $\mathbf{u}_e[i] \gets u_e^i$        
            \State $\mathbf{u}_a[i] \gets u_a^i$
        \EndFor
        \\
        \State BackPropagation($A$,  $\mathcal{J}(s_0, \mathbf{u}_a, \mathbf{u}_e)$) 
        \State Update($\theta_A$)
    \EndFor
    \\
    \For{$\text{iters}_D$} \Comment{Train $D$}
        \State $\mathbf{u}_e$=[\:]
        \State $\mathbf{u}_a$=[\:]
        \\
        \For{$i \gets 0 \dots H_\text{train}-1$}            
            \State $\textbf{z} \gets \mathcal{N}(0,1)$
            \State $u_e^i \gets A(\theta_A, \textbf{o}_e^i, \textbf{z})$
            \State $u_a^i \gets D(\theta_D, \textbf{o}_a^i)$
            \State $\mathbf{u}_e[i] \gets u_e^i$        
            \State $\mathbf{u}_a[i] \gets u_a^i$
        \EndFor
        \\
        \State BackPropagation($D$,  $\mathcal{J}(s_0, \mathbf{u}_a, \mathbf{u}_e)$) 
        \State Update($\theta_D$)      
    \EndFor
\EndProcedure
\end{algorithmic}
\end{algorithm}

\begin{algorithm}[H]
\begin{algorithmic}[1]
\Procedure{Test}{$\mathcal{M}$, $H_\text{test}$}
    \State $s_0 \gets$ GetState($\mathcal{M}$)
    \State $\xi$ := $[s_0]$
    \For{$i \gets 0 \dots H_\text{test}-1$}            
        \State $\mathbf{z} \gets \mathcal{N}(0,1)$
        \State $u_e^i \gets A(\theta_A, \textbf{o}_e^i, \textbf{z})$
        \State $u_a^i \gets D(\theta_D, \textbf{o}_a^i)$
        \\
        \State $\xi[i+1] \gets \psi(s_i, u_a^i, u_e^i, t_i)$
    \EndFor
    \State $\rho=R_\Phi(\xi)$
\EndProcedure
\end{algorithmic}
\end{algorithm}

\newpage

\section{Experimental settings}
\label{sec:settings}

\begin{table}[ht]
\caption{Training parameters and constraints for cart-pole problem with adversarial cart-track friction and moving target.
}
\centering
\def\arraystretch{1.5}
\begin{tabular}{c|c}
 \multicolumn{2}{c}{\textbf{Cart-pole with moving target}}\\
 \toprule
    Training iterations & 
    \def\arraystretch{1}\begin{tabular}{@{}c@{}} $500$ overall steps\\
    1 attacker step\\ 2 defender steps
    \end{tabular}\\
    \hline
    Time step size & $\Delta t = 0.05\;s$\\
    Time horizon & $H=40$\\
    Temporal depth & $h=10$\\
    \hline
    Initial cart position & $x_0 \sim \mathcal{U}(-1,1)\;m$\\
    Initial cart velocity & 
    $\dot x_0\sim \mathcal{U}(-1,1)$ $m/s$\\
    Initial pole angle & $\theta_0 \sim \mathcal{U}(-0.1,0.1)\;rad$\\
    Initial angular velocity &
    $\dot \theta_0\sim \mathcal{U}(-1,1)$ $m/s$\\ 
    \hline
    Position constraint & $x \in [-30\;m, 30\;m]$\\  
    Velocity constraint & $\dot x \in [-10\;m/s, 10\;m/s]$\\
    Angle constraint & $\theta\in[-1.5\;rad,1.5\;rad]$\\
    Friction constraint & $\mu\in[0,1]$\\
    Target position offset constraint & $\dot\epsilon\in[-5\;m/s,5\;m/s]$\\ 
    Robustness weight & $\alpha=0.4$\\
    \hline
    Attacker's architecture &
    \def\arraystretch{1}\begin{tabular}{@{}c@{}}1 layer with 10 neurons each\\ and Leaky ReLU activations\end{tabular}\\
    \rule{0pt}{4ex}
    Defender's architecture &
    \def\arraystretch{1}\begin{tabular}{@{}c@{}}2 layers with 10 neurons each\\ and Leaky ReLU activations\end{tabular}\\
    Noise space  & $\mathcal{Z}=\mathbb{R}^3$  \\
 \bottomrule
\end{tabular}
\label{tab:cartpole_params}
\end{table}


\begin{table}[ht]
\caption{Training parameters and constraints for car platooning problem with energy consumption requirements.
}
\centering
\def\arraystretch{1.5}
\begin{tabular}{c|c}
 \multicolumn{2}{c}{\textbf{Car platooning with energy consumption}}\\
 \toprule
    Training iterations & 
    \def\arraystretch{1}\begin{tabular}{@{}c@{}} $1000$ overall steps \\
    1 attacker step \\ 
    2 defender steps
    \end{tabular}\\
    \hline
    Time step size & $\Delta t = 0.05\;s$\\
    Time horizon & $H=40$\\
    Temporal depth & $h=10$\\
    \hline
    Initial distance & 
    \def\arraystretch{1}\begin{tabular}{@{}c@{}}$d_0\sim\mathcal{U}(2, 6)\;m$\\
    \end{tabular}\\
    Initial velocity & $\dot x_{l0}, \dot x_{f0}\sim\mathcal{U}(15,20)\;m/s$\\
    \hline
    Acceleration constraints & $\ddot x_l, \ddot x_f\in[-5\;m/s^2,5\;m/s^2]$\\
    Velocity constraints & $\dot x_l, \dot x_f\in[0\;m/s^2,37\;m/s^2]$ \\
    Torque constraints & $T_{m} \in [-T_{m,MAX}(\omega_m), T_{m,MAX}(\omega_m)]$\\
    Robustness weight & $\alpha=0.98$\\
    \hline
    Attacker's architecture &
    \def\arraystretch{1}\begin{tabular}{@{}c@{}}1 layer with 10 neurons each\\ and Leaky ReLU activations\end{tabular}\\
    \rule{0pt}{4ex}
    Defender's architecture &
    \def\arraystretch{1}\begin{tabular}{@{}c@{}}2 layers with 10 neurons each\\ and Leaky ReLU activations\end{tabular}\\
    Noise space  & $\mathcal{Z}=\mathbb{R}^2$  \\
\bottomrule
\end{tabular}

\label{tab:platooning_params}
\end{table}


\newpage
\section{STL semantics}\label{sec:semantics}
Assume signals $x_1[t],\ldots,x_n[t]$, then atomic predicates are of the form $\mu = f(x_1[t],\ldots,x_n[t])>0$.

\paragraph{Boolean semantics.}
The satisfaction of a formula $\varphi$ by a signal $\mathbf{x}=(x_1,\dots , x_n)$ at time $t$ is defined as:
\begin{itemize}
   \item $(\mathbf{x},t) \models \mu \iff f(x_1[t],\ldots,x_n[t])>0$;
   \item $(\mathbf{x},t) \models \varphi_1 \land\varphi_2 \iff (\mathbf{x},t) \models\varphi_1\land(\mathbf{x},t) \models\varphi_2$;
   \item $(\mathbf{x},t) \models\neg\varphi \iff \neg((\mathbf{x},t) \models\varphi))$;
   \item $(\mathbf{x},t) \models\varphi_1 U_{[a,b]}\varphi_2 \iff \exists t'\in [t+a,t+b] \mbox{ s.t. }\\ (\mathbf{x},t') \models\varphi_2\land \forall t''\in [t,t'], (\mathbf{x},t'') \models\varphi_1$.
\end{itemize}
\begin{itemize}
    \item Eventually:\\ $(\mathbf{x},t) \models \Diamond_{[a,b]}\varphi \iff \exists t'\in[t+a,t+b]  \mbox{ s.t. } (\mathbf{x},t') \models\varphi$;
    \item Globally:\\ $(\mathbf{x},t) \models \square_{[a,b]}\varphi \iff \forall t'\in[t+a,t+b]  \quad (\mathbf{x},t') \models\varphi$.
\end{itemize}

\paragraph{Quantitative semantics.}
The quantitative semantics, meaning the robustness, of a formula $\varphi$ is defined as a function $\rho^\varphi$:
\begin{itemize}
    \item $\rho^\mu (\mathbf{x},t) = f(x_1[t],\ldots,x_n[t])$;
    \item $\rho^{\neg\varphi}(\mathbf{x},t) = -\rho^{\varphi}(\mathbf{x},t)$;
    \item $\rho^{\varphi_1\land\varphi_2}(\mathbf{x},t) = \min (\rho^{\varphi_1}(\mathbf{x},t),\rho^{\varphi_2}(\mathbf{x},t))$;
    \item $\rho^{\varphi_1 U_{[a,b]}\varphi_2}(\mathbf{x},t) =\\ \underset{\tau\in [t+a,t+b]}{\sup}\left(\min\left(\rho^{\varphi_2}(\mathbf{x},\tau), \underset{s\in [t,\tau]}{\inf}\rho^{\varphi_1}(\mathbf{x},s)\right)\right) $.
\end{itemize}
The sign indicates the satisfaction status: 
\begin{itemize}
    \item[-] $\rho^\varphi(\mathbf{x},t)>0\iff (\mathbf{x},t)\models\varphi$;
    \item[-] $\rho^\varphi(\mathbf{x},t)<0\iff (\mathbf{x},t)\not\models\varphi$.
\end{itemize}


\newpage
\section{Additional plots}\label{sec:additional_plots}

\begin{figure}[!hb]
    \centering
	\includegraphics[width=8.5cm, keepaspectratio]{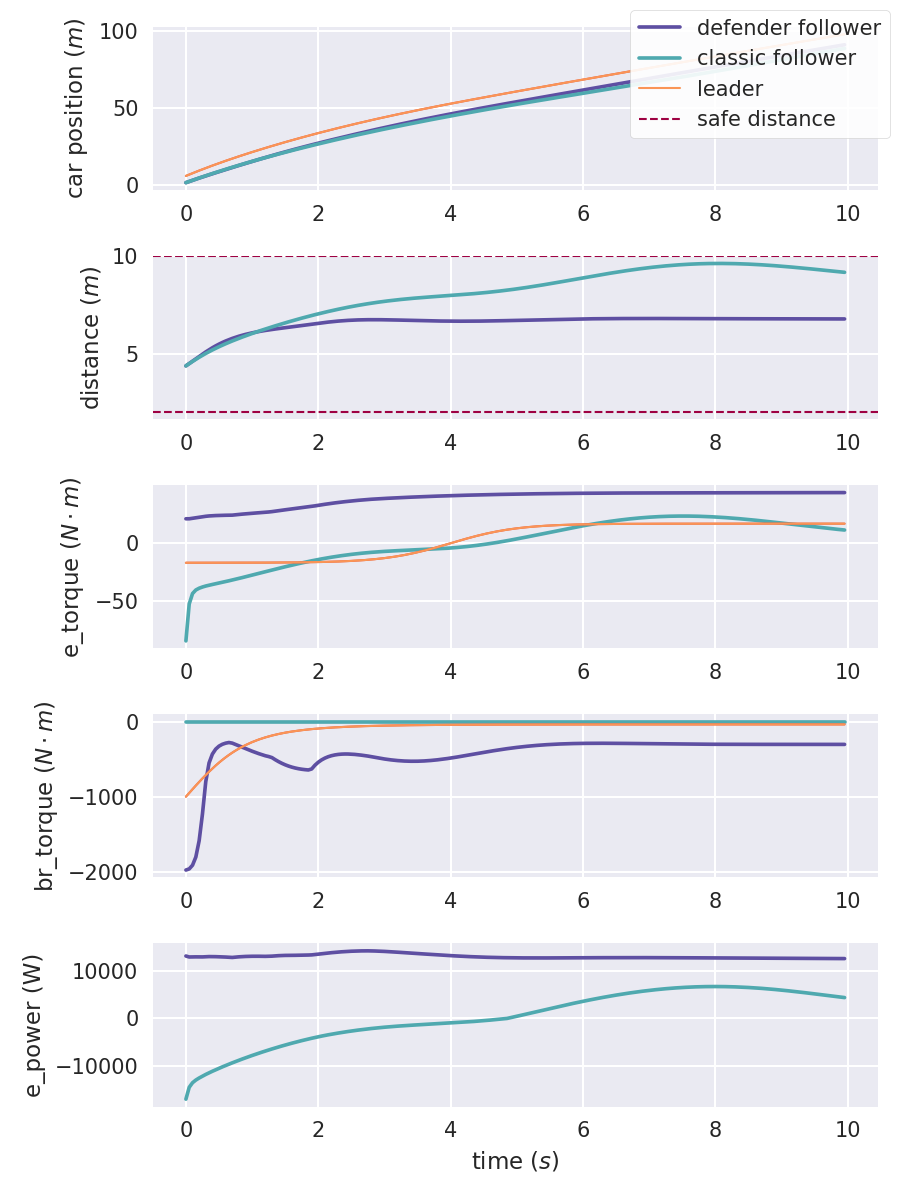}
	\caption{Two different followers against the same leader: \emph{defender} car acts according to the Defender network, \emph{classic} car is managed by a classical controller. In both cases the evolution begins with the same initial configuration: distance between cars $d=2.9356\;m$, leader velocity $v_l=17.3107\;m/s$, follower velocity $v_f=16.3543\;m/s$.	The time horizon is $H=200$ and the step width is $\Delta t=0.05\,s$. 
	}
	\label{fig:platooning_pulse_evolution}
\end{figure}

\begin{figure}[!hbt]
\centering	 
\includegraphics[width = \columnwidth]{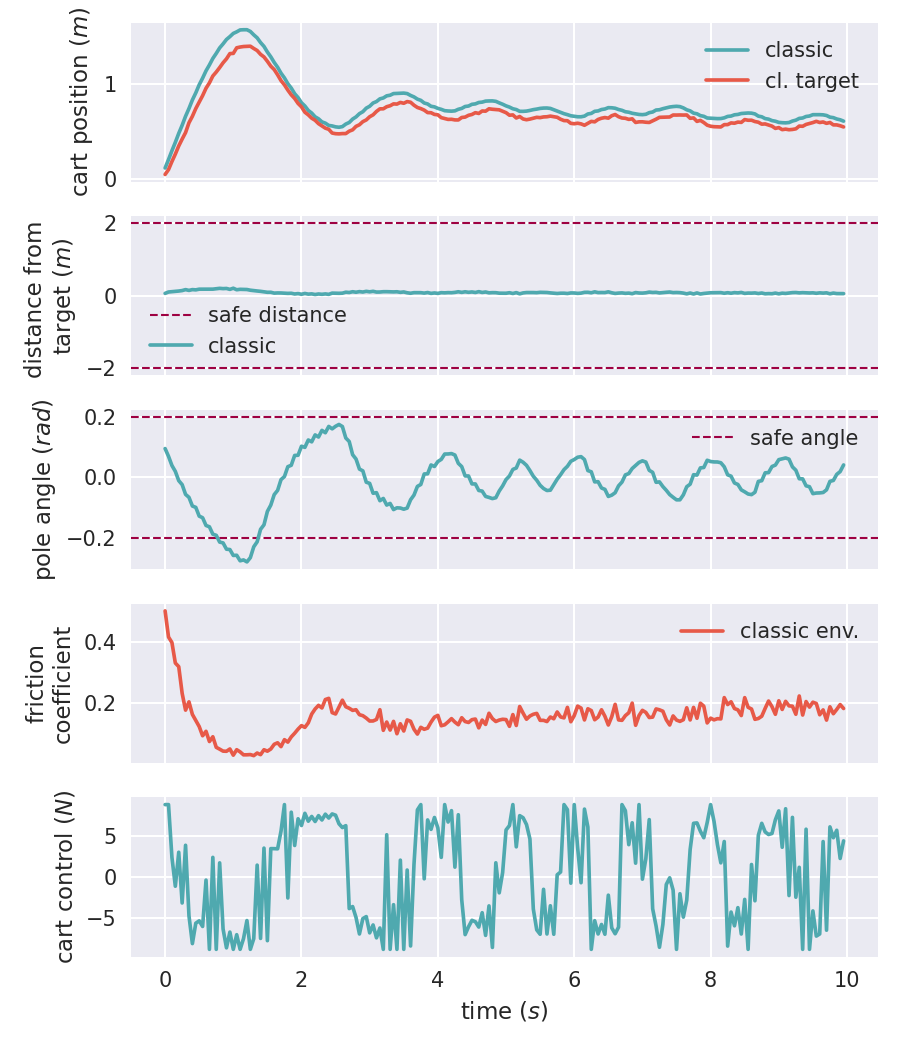}
\caption{Evolution of the system for cart-pole balancing problem with moving target, using a classical controller for the cart against the Attacker network, which generates track-cart friction coefficient and target.
Initial configuration: cart position ${x=0.7835}\;m$, cart velocity ${\dot x=0.9550}\;m/s$, pole angle ${\theta=0.0715}\;rad$, pole angular velocity ${\dot\theta=0.8155}\;rad/s$.}
\end{figure}

\begin{figure}[!hbt]
\centering	 
\includegraphics[width = 0.9\columnwidth]{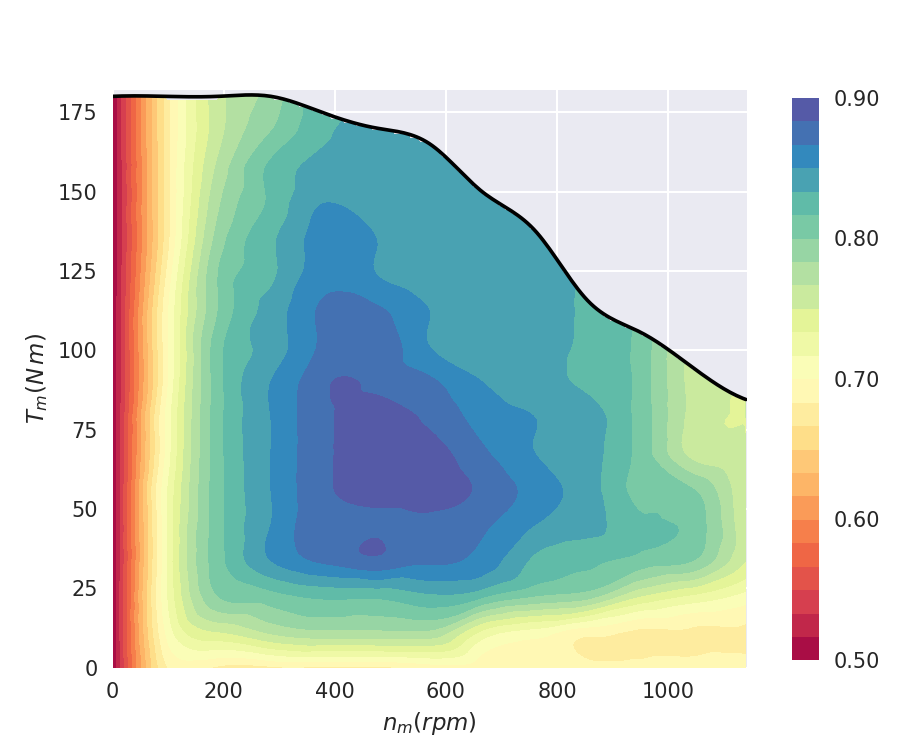}
\caption{Powertrain efficiency map, speed $n_m$ expressed in $rpm$. Efficiency values are assumed symmetrical in the negative torque case. For positive values, electric torque is bounded by the maximum torque value $T_m^{MAX}(\omega_m)=180 \,Nm$, speed is bounded by the maximum value $n_m^{MAX} = 1140 \,rpm$.}
\label{fig:eff_map}
\end{figure}

\end{document}